\documentclass[amssymb, 
amsmath, 
amsfonts,
reprint,
superscriptaddress,
longbibliography,
physrev, 
aps]{revtex4-2}

\usepackage[utf8]{inputenc}
\usepackage{graphicx}
\usepackage{dcolumn}
\usepackage[T1]{fontenc}
\usepackage{tabularx}
\usepackage{enumitem}
\usepackage{currvita}
\usepackage{filecontents}

\usepackage[mathlines]{lineno}
\usepackage{graphicx}
\usepackage[margin=1in]{geometry} 
\usepackage{graphicx}
\usepackage[table]{xcolor}

\usepackage[hidelinks]{hyperref}
\definecolor{color1}{rgb}{0,0.25,0.70}
\hypersetup{colorlinks=true, 
    linkcolor={color1},
    citecolor={color1}, 
    urlcolor={color1}
}

\newcommand{\closedm}[1]{\left[ #1 \right]}
\newcommand{\closeds}[1]{\left( #1 \right)}

\newcommand{\ket}[1]{| #1 \rangle}
\newcommand{\bra}[1]{\langle #1 |}
\newcommand{\kett}[1]{| #1 \rangle \px{-2} \rangle}
\newcommand{\bbra}[1]{\langle \px{-2} \langle #1 |}
\newcommand{\closedee}[1]
{\langle \px{-2} \langle #1 \rangle \px{-2} \rangle}
\newcommand{\px}[1]{\hspace{#1 pt}}

\begin{document}

\title{Electronic superradiance mediated by nuclear dynamics\\}

\author{Xuecheng Tao}
\affiliation{Division of Physical Sciences, College of Letters and Science, University of California Los Angeles, Los Angeles, California 90095, USA}
\author{John P. Philbin}
\affiliation{Division of Physical Sciences, College of Letters and Science, University of California Los Angeles, Los Angeles, California 90095, USA}
\author{Prineha Narang}
\email{email: prineha@ucla.edu}
\affiliation{Division of Physical Sciences, College of Letters and Science, University of California Los Angeles, Los Angeles, California 90095, USA}
\affiliation{Electrical and Computer Engineering Department, University of California, Los Angeles, California, 90095, USA}

\begin{abstract}
Superradiance, in which the collective behavior of emitters can generate enhanced radiative decay, was first predicted by a model, now known as the Dicke model, that contains a collection of two-level systems (the emitters) all interacting with the same photonic mode.
In this article, we extend the original Dicke model to elucidate the influence of nuclear motion on superradiant emission.
Our dynamical simulations of the combined electronic, nuclear, and photonic system reveal a new time scale attributed to the population leakage of the dark, subradiant states.
Furthermore, this dark state emission pathway can be controlled by tuning the nuclear potential energy landscape.
These findings impact how superradiant states and molecular degrees of freedom can be leveraged and utilized in quantum optical systems.
\end{abstract}

\maketitle
\clearpage

{\it Introduction.---}
Superradiance, 
a concept originally proposed by Dicke in 1954 \cite{dicke_coherence_1954},
describes a scenario in which enhanced radiative decay is obtained from an ensemble of emitters interacting cooperatively with the same photonic mode of the radiation field.
Dicke superradiance (and superfluorescence) \cite{dicke_coherence_1954, gross_superradiance_1982, leonardi_dicke_1986, lin_superradiance_2012} 
has been experimentally observed in a number of platforms including
atomic and molecular, nanomaterials, and solid-state systems \cite{
skribanowitz_observation_1973, malcuit_transition_1987, scheibner_superradiance_2007, 
timothy_noe_ii_giant_2012, dai_observation_2011,  
miyajima_superfluorescent_2009, auerbach_super-radiant_2011, vasil2009femtosecond, mlynek_observation_2014, de_oliveira_single-photon_2014, chen_experimental_2018,
bradac_room-temperature_2017, cong_dicke_2016, raino_superfluorescence_2018, 
biliroglu_room-temperature_2022, huang_room-temperature_2022}.
Furthermore, there has been revived interest in superradiance for its connections to
the laser and electron beam implementations \cite{kocharovsky_superradiance_2017, gover_superradiant_2019}, 
entangled states manipulation \cite{ficek_entangled_2002}, and
quantum simulators \cite{lamata_digital-analog_2018, wang_controllable_2020}
beyond the traditional quantum optics context.

A key element of the cooperative behavior in Dicke's modeling of superradiance 
is that all emitter dipoles interact uniformly with the commonly experienced radiation mode. \cite{dicke_coherence_1954}
However, when the experiments
are performed with emitters that are more complex than structureless atoms and at finite temperatures, 
the motions of nuclear degrees of freedom break the uniformity in the emitter behaviors. 
For example, one type of non-uniformity, onsite transition energy disorder,
has been shown to influence radiative lifetimes of superradiant systems \cite{philbin_room_2021, blach_superradiance_2022}. 
To this end, several studies \cite{cederbaum_cooperative_2022, bustamante2022tailoring, chen_interplay_2022, damanet2016cooperative, damanet2016master, facchinetti2016storing, sandner2011spatial, sauiko2002theory} have explored the significance of incorporating individual nuclear motions into the theoretical analysis of collective emissions from diverse perspectives.
Cederbaum\cite{cederbaum_cooperative_2022} showed that cooperative effects are impacted by the inclusion of individual vibrational coordinates in the modeling
of collective states, even under the rather strict assumption of uniformity of the coordinates across the sites required to perform analytical analysis.
Chen et al. \cite{chen_interplay_2022} modelled the environmental disorder as a stochastic modulation of the excitation energies and found that this disorder can effectively restore the coherence in the emitter ensemble.
Especially, it is worth mentioning that Bustamante et al. \cite{bustamante2022tailoring} found that the accumulation or release of the energy in the collective state depends strongly on the distance and orientation of the molecules in a linear array, using a realistic potential from quantum chemistry calculations.

In this article, we enhance the familiar Dicke model by adding nuclear degrees of freedom to the electronic emitters and investigate their influence on the electronic superradiant dynamics.
We build a dynamics model to simulate the concerted electronic-photonic-nuclear interactions
to determine how the emission rates can be modified by nuclear dynamics. 
We aim at constructing a simulation model with minimal computational complexity, allowing a focused examination of the effects of nuclear vibrations.
Specifically, the emission signals are explicitly modeled with the system evolution, in contrast to the 
matter-only description \cite{clemens_collective_2003, damanet_atom-only_2019, masson_universality_2022, sierra_dicke_2022} in the superradiance literature. 
We observe that the nuclear dynamics generate dynamic disorder and leads to an emission pathway for traditionally dark, subradiant states.

\begin{figure*}[!t]
\includegraphics[width=2\columnwidth]{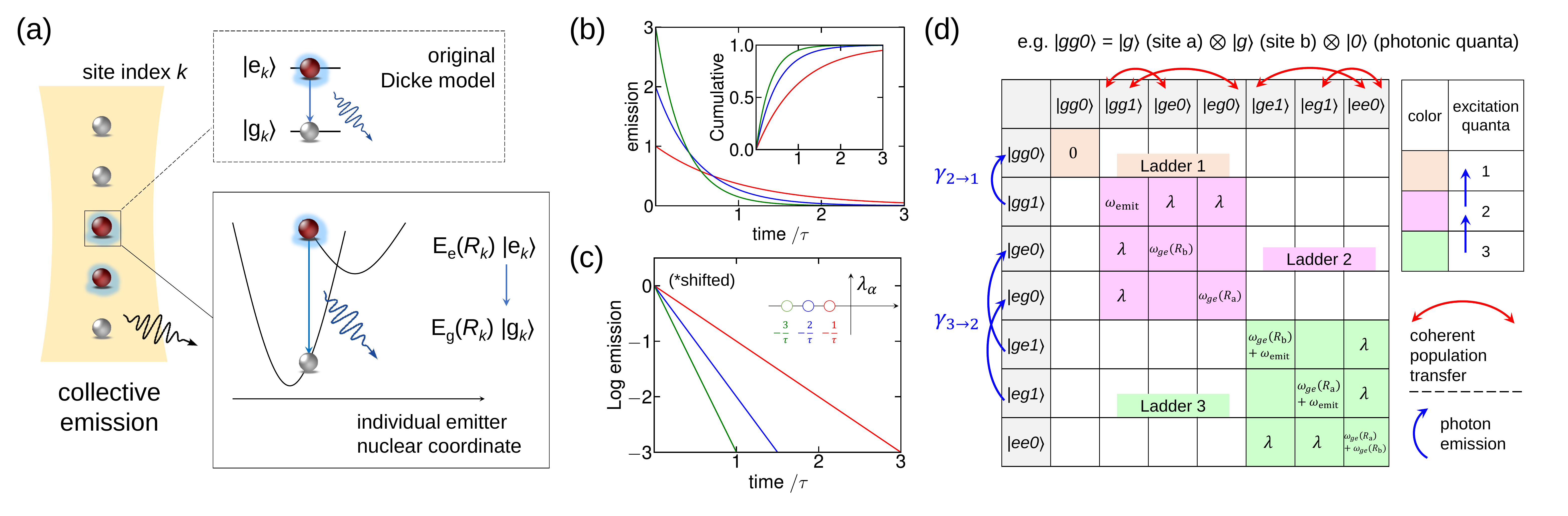}
\caption{\label{fig:scheme}
\textbf{Dicke ladder for the superradiance emission, revisited with the explicit emission model.}
(a) Collective emission from an ensemble of interacting emitters. 
The emitters are considered as structureless two-level systems (dashed box), 
or with the nuclear degrees of freedom (solid box). 
(b) The instantaneous and the cumulative (inset) emitted photon count as detected emission signals, for the cases of 1-(red), 2-(blue), 3-sites (green) single photon superradiance. 
(c) Log scale plot of the emission signal in (b). The intercepts are shifted for a better comparison of the slopes.
Note that the slopes correspond to the emission rates that Dicke predicted in Ref.~\cite{dicke_coherence_1954}. The inset plots the emission rates as the real negative eigenvalues from the spectral decomposition of Liouvillian operator.
(d) Illustrations of the Hamiltonian and the primary dynamical processes in the explicit emission model, 
exemplified with the two-site collective emission example. }
\end{figure*}

{\it Model.---}
Dicke first investigated \cite{dicke_coherence_1954} the collective emission 
of structureless emitters with no spatial overlap with each other 
and that only interacts directly with the photonic mode.
The Tavis-Cumming Hamiltonian \cite{tavis_exact_1968} presents the simplest framework to analyze this scenario of collective emission (set $\hbar=1$), 
$H = \sum_{k=1}^N \left[ \frac12 \omega_{ge} \sigma^{(k)}_z 
      + \omega_{\rm emit} a^{\dagger} a
      + \lambda (\sigma^{(k)}_+ a + \sigma^{(k)}_- a^{\dagger}) \right]$,
$\sigma$ are the Pauli operators for the emitters, 
i.e. $\sigma_- \ket{e} = \ket{g}$ represents an emission process that happened with the release of a photon, $a^{\dagger}$ (see Fig.~\ref{fig:scheme}a),
and $k$ indexes the emitter sites; 
$\omega_{ge} = \varepsilon_e - \varepsilon_g$ is the energy splitting, 
$\omega_{\rm emit}$ is the emission frequency, and $g$ is the light-matter coupling strength.
We focus on the condition of (near) free-space spontaneous emission, 
where $\omega_{ge} - \omega_{\rm emit} \ll 1$, and $\lambda \ll 1$.
In the above Hamiltonian, we have expressed the light-matter coupling 
under the rotating-wave approximation\cite{jaynes_comparison_1963}.

The simple, analytical but profound results derived by Dicke, despite the complexity of the many-body Hamiltonian,
effectively exploit the assumption that all individual emitters interact with the radioactive field in an {\it identical} manner.
As a result, one only needs the collective operators,
$\sum_k \sigma^{(k)}_z$ and 
$ \lambda \left( \sum_k  \sigma^{(k)}_+ a + \sum_k \sigma^{(k)}_- a^{\dagger} \right) $,
instead of the $N$ individual operators,
to derive the selection rules and the rates for the collective emission. \cite{dicke_coherence_1954}
The emission rates are obtained as proportionally to the square of the matrix elements of the transition operators,
following Fermi's golden rule.
It is predicted that the emission rate can be accelerated with an increasing number of emitters.
For example, when the system is prepared in a singly-excited collective state, in which the excitation is shared symmetrically among the $N$ emitters, it is predicted that the emission rate increases linearly with $N$. This is known often referred to as ``single-photon superradiance''.
Furthermore, an enhancement of the emission intensity that is proportional to $N^2$ can be obtained by preparing the system with all emitters excited and letting the system radiate half of the stored photons, creating a ``superradiant burst''.

To illustrate how nuclear motions intervene with electronic emission, it becomes necessary to move beyond the Fermi's-golden-rule description of emission rates to resolve competing time scales in the superradiance emission.
To this end,
we augment the original Dicke model with nuclear and photonic degrees of freedom (DOFs) and 
the emission signal is explicitly computed as a function of the system density matrix.
Thus our model is referred to as the explicit emission model, while the conventional matter-only description is referred to as the implicit emission models hereafter.
We closely follow the setup in Dicke's model,
and add the nuclear DOFs dependency in the matter Hamiltonian in an extended state representation
\begin{align} \label{eq:hamiltonian}
      H_{\rm M} = & \sum_k P_k^2/2m_k \px{2} \mathbb {I} \nonumber \\
      & + \sum_k \left[ \varepsilon_g(R_k) \ket{g}\bra{g}_k  + \varepsilon_e(R_k) \ket{e}\bra{e}_k \right]
\end{align}
When the nuclear DOFs are frozen ($P_k=0, \px{2} R_k=R_0$ for all $k$), 
the matter Hamiltonian recovers the familiar two-level description 
$\sum_k \frac12 \omega_{ge} \sigma^{(k)}_z$ in the Dicke model, up to a constant energy shift.
We have the same light-matter interaction Hamiltonian as
\begin{align}
H_{\rm L-M} = & \sum_k \lambda \closedm{ \sigma^{(k)}_- a^\dagger + \sigma^{(k)}_+ a }.
\end{align}
In the following, we set $\lambda$, the coupling strength, to be the same value for all sites and 
focus on the effects non-uniform nuclear dynamics ($R_k$ and $P_k$) have on the emission rates.

We then perform time-dependent simulations of the emission dynamics generated by the above Hamiltonian,
with simple choices of the nuclear potential energy surfaces, $\varepsilon_g(R)$ and $\varepsilon_e(R)$, to elucidate how the emission dynamics can be modified by nuclear motions.
In the simulations, a time-scale separation of the fast photon-electron motion and the slower nuclear motion is employed, i.e. 
$\rho = \rho_{\rm nuc}(R) \otimes \rho_{\rm pho-ele}$;
the first term on the right-hand side, the nuclear density matrix, is expressed on a phase-space grid and the second term, the photon-electron density matrix is expressed in the state representation,
a tensor product basis of electronic states and photonic Fock states. 
The time evolution of the nuclear-photonic-electronic density matrix is described with the Markovian master equation,
\begin{align} \label{eq:lindblad}
    \dot{\rho}(t) = \mathcal{L} \rho 
    = -i[H, \rho]  + \gamma  \mathcal{D}(a).
\end{align}
In Eq.~\ref{eq:lindblad}, the first term on the r.h.s. is the unitary Von-Neumann evolution, 
and the second term, a Lindblad term \cite{lindblad_generators_1976, gorini_completely_2008, manzano_short_2020}, 
$ \mathcal{D}(a) = 
    a \rho a^{\dagger} -
    \closeds{a^{\dagger} a \rho + \rho a^{\dagger} a} / 2 $
accounts for the dissipative photon emission with $\gamma$ the loss strength.
The emission rate (the broadening of the emission peaks is not considered here)
are then explicitly modeled with the dissipative dynamics, 
is both linear to the photon loss strength and the populations of the emissive states.

{\it Results and discussion.---}
We open the discussion of results by first convincing the readers that our explicit emission model reproduces the 
same physics as the conventional implicit emission modeling.
In this subsection, nuclear DOFs are frozen ($R_k=P_k=0$) as in the original Dicke model.
In the elementary example of a single-site emission,
the space of the photon-electronic density matrix is spanned by the basis $\{ \ket{g,0}, \ket{g,1}, \ket{e,0} \}$
(e.g. $\ket{g,0}$=$\ket{g}$ (site a) $\otimes \ket{0}$ (photonic number)),
Note that as Dicke did, we restrain ourselves in the single-excitation photonic subspace.
Given the free-space emission condition, $\lambda / \gamma \ll 1$, 
an analytical solution can be derived with the steady-state approach \cite{houston2012chemical}
for the state populations throughout the dissipative process. 
When the system is initialized from the emissive state,
${\rho}_{\rm pho-ele}(0) = \ket{e,0} \bra{e,0}$,
the state populations are 
\begin{subequations} \label{eq:1site_pop}
\begin{align} 
    \mathbb{P}_{\ket{g,0}}(t) &= 1 - e^{-t / \tau}, \px{10}
    \mathbb{P}_{\ket{e,0}}(t) = e^{-t / \tau},
\end{align}
\end{subequations}
where $\px{6} \tau = \gamma / 4 \lambda^2$. 
As a consequence, we have an exponentially shaped emission signal after the initial induction period and the emission rate is $1/\tau$.
The system must reach an intermediate, photonically excited state $\ket{g,1}$ 
to complete the emission process $\ket{g,1} \to \ket{g,0}$ in our modeling, 
and it is effectively equivalent to a direct $\ket{e} \to \ket{g}$ process in the implicit modeling when the photon decay to the environment is fast compared to the coherent population transfer dynamics  (i.e. when $\lambda / \gamma \ll 1$).

The emission rates can also be accessed by analyzing the structure of the Liouvillian superoperator with spectral decomposition, which has a more favorable computational scaling than the steady-state solution.
With Eq.~\ref{eq:lindblad}, the evolution of the system can be expressed \cite{albert_symmetries_2014} by vectorizing the density matrix in the Fock-Liouvillian space, namely,
\begin{subequations}  \label{eq:spectral_evolution}
\begin{align}
    \kett{\rho(t)} &= e^{\ell t} \kett{\rho(t=0)} 
    = \sum_{\alpha} e^{\lambda_\alpha t} \px{2} Q_\alpha     \px{2} \kett{R_\alpha}, \\
    Q_\alpha &= \sum_{\beta}     \frac1{\closedee{L_\alpha | R_\beta }}
    \closedee{L_\beta | \rho(0)}.
\end{align}
\end{subequations}
In the above expression, we have used the symbol $\kett{\cdot}$ for the vectorized space and $\ell$ for the matricized Liouvillian superoperator. 
$\bbra{L_\alpha}$ and $\kett{R_\alpha}$ are the left and right eigenvectors of $\ell$, respectively.
Fig.~\ref{fig:scheme}b plots the instantaneous dissipated photon count for $N=1,2,3$ 
single photon superradiance emission initialized in the symmetric Dicke state.
Fig.~\ref{fig:scheme}c takes the logarithm of the emission signal in (b), 
and the slopes of these curves, respectively, give the emission rates of $1/\tau$, $2/\tau$ and $3/\tau$. This linear increase of the emission rates with the number of emitter sites involved in the radiative decay is consistent with previous works. \cite{dicke_coherence_1954, scully_super_2009}
As shown with Eq.~\ref{eq:spectral_evolution}, the emission time scales are encoded in the real negative eigenvalues in the spectral decomposition of the Liouvillan operator (inset of Fig.~\ref{fig:scheme}c).

The full ``Dicke ladder'' of the two-site collective emission can also be reproduced with our explicit emission model, as shown schematically in Fig.~\ref{fig:scheme}d.
The square matrix in panel Fig.~\ref{fig:scheme}d illustrates the photon-electronic Hamiltonian.
Note that the Tavis-Cumming Hamiltonian is block diagonal in the photon-electronic state basis and the unitary evolution under the Tavis-Cumming Hamiltonian preserves the total excitation number $\langle a^\dagger a + \sum_k \sigma_z^{(k)} / 2 \rangle$ (see the colored subspaces).
The red arrows describe the coherent population transfer within each excitation subspace and the blue arrows describe the dissipative photon loss that moves the system down the ladder to the below excitation subspace with one less photon.
This dissipative photon loss is the source of emission signals in our simulation model.
From the spectral decomposition of the Liouvillian,
we obtain the stationary states as the eigenvectors associated with eigenvalues equal to zero. 
We find that both the fully dissipated state $\ket{g...g,0}$ and the dark (i.e. subradiant) state $( \ket{eg,0} - \ket{ge,0}) / \sqrt{2}$ in the $N=2$ case) survive in the infinite time limit.
We also obtain the real negative eigenvalues that account for the dissipative photon loss, and the emission time scales.
Furthermore, by switching on and off the dissipation channels separately (e.g. set $\gamma_{2 \to 1}$=0, $\gamma_{3 \to 2}=\gamma_0$), we observe the single photon emission processes. 
The resulting eigenvalues correspond to the spontaneous emission rates predicted by Dicke, and the eigenvectors with non-zero eigenvalues are interpreted as emission pathways. 
It is worth mentioning that we observe in the numerical simulation that it is not able to resolve all the emission pathways when the Liouvillian is not diagonizable. The observation motivates the numerical simulation strategy for the time evolution of the density matrix in the next section.

Next we establish connections between the implicit and explicit approaches to modeling the emission process.
An emission event in the matter-only model
translates into a two-step process in the explicit emission model involving a photonically excited intermediate state. 
Both approaches can describe the same underlying physical process when the photon dissipation rate is large.
Although including photonic degrees of freedom in a free-space emission simulation seems redundant at first glance, there are at least three unique advantages of this approach.
Firstly, the explicit modeling encompasses the ability to resolve multiple competing time scales within the radiative process, which is crucial for our purposes, to identify the effects of nuclear DOFs on electronic emission.
Secondly, the explicit modeling offers the advantage of discerning between radiative and non-radiative transitions, and finally, the explicit modeling avoids the perturbative treatment of the light-matter interaction and explicitly incorporates the evolution of photonic DOFs, making it useful in studies where the characteristics of the radiative field are of vital interest \cite{sanvitto_road_2016, fribeiro_polariton_2018, flick_strong_2018, tibben_molecular_2023}. 

\begin{figure*}[!t]
\includegraphics[width= 2\columnwidth]{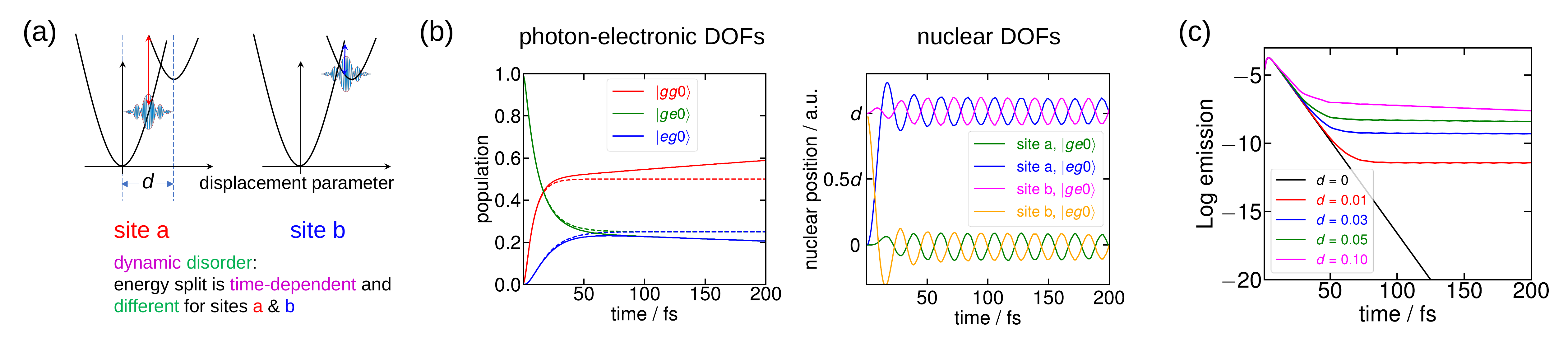}
\caption{\label{fig:dynamic_disorder}
\textbf{Dynamic disorder originated from nuclear vibrations
mediates the electronic emission in two-site superradiance.}
(a) Schematic plot for the dynamic disorder in the two-site collective emission model.
The equilibrium positions of the nuclei exhibit a displacement, $d$, between the emissive state ($\ket{e}$)
and the emitted state ($\ket{g}$). 
(b) State populations (left) and the nuclear positions (right) of the emission system as a function of time evolution.
The nuclear positions are obtained as expectation values averaged over the nuclear wavepackets.
The emissions are influenced by nuclear motions (solid, $d=0.05$), with those obtained without the dynamic disorder (dashed, $d=0$) included as a guide to the eyes. 
(c) Instantaneous emission signals in the logarithm scale for emission models with increasing degrees of dynamic disorder.
}
\end{figure*}

With the explicit emission model validated, we now investigate the influence of nuclear vibrations in superradiance with numerical approaches.
Simple but sufficient to illustrate, we model the nuclear DOFs with the displaced harmonic oscillator approximation---the emissive and the emitted states are assumed to share equal vibrational frequencies but differ in their equilibrium positions (see Fig.~\ref{fig:dynamic_disorder}a).
The displacement parameter, $d$, is specific to the electronic and molecular structure of the emitter and impacts the spectroscopy and utilities of molecules in various ways.
For example, small $d$ values are ideal when using emitters as optical cycling centers to try and laser cool molecules \cite{dickerson_single_2023}. 
Larger disparities in the nuclear potential energy landscapes between the emissive and emitted states, as quantified by a higher displacement parameter, lead to more dynamic disorder, and are expected to break the uniformity of the emitter behavior to a larger extent.
To quantify the impact of dynamic disorder on superradiant emission, we perform the numerical simulations 
with the rapid photon-electronic DOFs described in the truncated photon-electronic basis,
and with the slower nuclear DOFs described as wavepackets on a phase-space grid. 
Model parameters and computational details are reported in the Appendix.

Fig.~\ref{fig:dynamic_disorder}b presents the time evolution of the system for both photon-electronic and nuclear degrees of freedom in a two-site single photon superradiance emission scenario.
In the simulation, the system is initialized as an electronically emissive state 
$\ket{ge,0}=\ket{g} $ (site a) $\otimes \ket{e} $ (site b) $\otimes  \ket{0} $ (photonic number),
and the nuclear wavepacket is initialized as the vibrational ground-state wavefunction,
centered around the equilibrium positions for the emitted (site a) and emissive (site b) states, respectively.
In the original Dicke model where nuclear DOFs are fixed (equivalent to model $d=0$, dashed, left panel), 
the cumulative emission signal (equal to $\mathbb{P}_{\ket{gg,0}}$ in this case) initially accumulates at twice the rate of that in a single-site emission; 
afterwards (at $t> \tau \sim$ $16$~fs), the cumulative emission signal reaches a photon count of $0.5$ (dashed red, left) and ceases, with the remaining photon stored in the collective subradiant state that are devoid of radiative decay.
However, as shown in Fig.~\ref{fig:dynamic_disorder}b, the inclusion of nuclear motions introduces a new emission time scale.
The photon-electronic process drives the nuclear wavepacket to move and oscillate around the new equilibrium position (solid blue and yellow, right panel).
The frequency of the wavepacket oscillation remains consistent across models with varying values of the displacement parameters, but the amplitude is larger in models with larger displacements.
Those nuclear motions introduce a feedback loop and impact the electronic emission process by generating dynamic disorder among different sites. 
In the physical picture where the nuclear and photon-electronic time scales are separated, the traditionally dark, subradiant state at the previous time snapshot has a small but non-zero overlap with the subradiant state at the next snapshot because of the vibrational motion between the two timesnaps.
As a result, the subradiant state is no longer perfectly dark. 
It therefore leads to a leakage of the photonic population stored in the dark state to the bright state, inducing dark-state emission (solid blue and green, left panel).
Note that when the disorder is modeled as a constant time-independent energy shift among the emitter sites this additional emission time scale would not be observable due to the lack of overlap between the subradiant and the bright states.

In Fig.~\ref{fig:dynamic_disorder}c, we quantify the timescales that are associated with the emission from the subradiant state population by plotting the logarithm scale of the instantaneous emission signals.
Compared to those reported in Fig.~\ref{fig:scheme}c, we identify clearly the additional time scale that appears after the superradiance time scale $\tau$.
For the models studied here,
it is observed that the emission rate of the photonic population in the subradiant state escalates with an increasing $d$.
In addition, the transition from superradiant emission to the slower emission appears earlier with larger $d$, or with larger dynamic disorder.
Since the collective dark states have been regarded as promising candidates for sensing \cite{shugayev2022hong} and information storage purposes \cite{scully_single_2015}, our results underscore the potential to further harness these states as long-lived communication channels by controlling the potential experienced by nuclear DOFs.

We conclude the discussion by outlining several experimental setups where connections with the theoretical findings could be established.
For instance, radiative lifetime, $\tau$, can be measured (e.g. with a similar setup in Ref. \cite{zhu2022functionalizing}) and compared to theoretical predictions for single molecule superradiance containing multiple, spatially separated optical cycling centers \cite{dickerson_single_2023}. 
Moreover, because the displacement parameter, $d$, is an intrinsic molecular vibrational property, measurements on superradiant molecules with different functional groups (i.e. different $d$s considered in this study) can be conducted to verify the  discoveries in this study.
In the experimental platforms based on trapped ions \cite{devoe1996observation} or arrays of coupled quantum emitters \cite{rui2020subradiant, lange2024superradiant}, 
the ability of high-precision control contributes to an easier observation of the effects of nuclear motions considered here. For instance, Ref.~\onlinecite{khan2023laser} proposed a laser cooling scheme for trapped ion system, leveraging the dark state emission pathways in particular.
Nanocrystals \cite{bradac_room-temperature_2017} and colloidal nanocuboids \cite{philbin_room_2021} also offer a pristine, microscopic solid-state platform to showcase the effects of the phononic dynamics, especially in Ref.~\onlinecite{philbin_room_2021}, effects from one form of the disorder was already observed.

{\it Conclusion.---}
In summary, we report that nuclear vibrations are able to mediate electronic superradiance by offering an additional emission pathway that is associated with emission from otherwise completely dark, subradiant states.
The investigation is performed with an extended Dicke collective emission model, 
where the dynamics of nuclear, electronic, and photonic degrees of freedom are explicitly described with quantum mechanics.
While the simulation currently utilizes a model potential energy surface (PES), it can readily integrate PES from quantum chemistry calculations for {\it ab initio} investigations, opening the possibilities to explore molecule-specific properties that influence the collective light-matter processes.
Other future directions include semiclassical modeling of the emission process
with trajectory-based methods \cite{li_quasiclassical_2020, koessler_incorporating_2022}
to avoid the costly computational effort with grid-based wavepacket simulations. 

\section*{Acknowledgements}
X.T. acknowledges Dr. Tomislav Begu\v{s}i\'c, Linqing Peng, Jack Diab and Dr. Davis Welakuh for helpful discussions. 
This work is supported by the Office of Naval Research (ONR) MURI Program under Grant Number ONR N00014-21-1-253. 
This research used resources of the National Energy Research Scientific Computing Center, a DOE Office of Science User Facility supported by the Office of Science of the U.S. Department of Energy under Contract No. DE-AC02-05CH11231 using NERSC award BES-ERCAP0025026. P.N. gratefully acknowledges support from the John Simon Guggenheim Memorial Foundation (Guggenheim Fellowship) as well as support from a NSF CAREER Award under Grant No. NSF-ECCS-1944085.

\begin{appendix}
\section*{Appendix: Computational Details}
We outline the numerical details regarding the coupled light-matter system in the section. 
First, the nuclear potentials are described with the displaced harmonic oscillator models, 
i.e. $\varepsilon_g(R) = \frac12 \omega_g R^2$, 
and $\varepsilon_e(R) = \frac12 \omega_e (R-d)^2 + s$
with $\omega_g = \omega_e = 1500$cm$^{-1}$, $s = 2$ eV, and $d$ are reported separately for models with different degrees of dynamic disorder in the main text. 
The nuclear masses $m=1$ amu for all the emitter sites.
For light-matter coupling parameters, we choose $\omega_{\rm emit} = s$ and ignore the emission peak broadening,
and $\lambda = 0.1$ eV, $\gamma=1$ eV such that the free-space emission condition $\gamma/\lambda \gg 1$ is met.

We perform the nuclear wavepacket simulation with the split-operator Fourier transform algorithm \cite{tannor_introduction_2007}, where the kinetic and potential operators for the nuclear are propagated in the momentum and coordinate space, respectively. 
The unitary evolution of photon-electronic states, as well as the dissipative evolution, are simulated with 4th-order Runge–Kutta methods \cite{press_numerical_2007}. 
In the simulations, the nuclear wavepacket is represented on a uniform real-space grid of 32 grid points at each dimension per emitter site, with the grid spanning Cartesian coordinates $[-2, 2)$ in atomic units. The timestep is $1.25$ fs.
\end{appendix}

\clearpage
%

\nocite{*}

\end{document}